\documentclass[epj]{svjour}
\usepackage{graphicx}
\begin{document}
\title{Transition from coherence to bistability in a model of financial markets}
\author{R. D'Hulst \and G. J. Rodgers
}                     
%
%
\institute{Department of Mathematical Sciences, Brunel University, Uxbridge, Middlesex UB8 3PH, UK}
\date{Received: date / Revised version: date}
%
\abstract{
We present a model describing the competition between information transmission and decision making in financial markets. The solution of this simple model is recalled, and possible variations discussed. It is shown numerically that despite its simplicity, it can mimic a size effect comparable to a crash. Two extensions of this model are presented that allow to simulate the demand process. One of these extensions has a coherent stable equilibrium and is self-organized, while the other has a bistable equilibrium, with a spontaneous segregation of the population of agents. A new model is introduced to generate a transition between those two equilibriums. We show that the coherent state is dominant up to an equal mixing of the two extensions. We focuss our attention on the microscopic structure of the investment rate, which is the main parameter of the original model. A constant investment rate seems to be a very good approximation.
\PACS{
      {02.50.Le}{Decision theory and game theory}   \and
      {02.50.Ng}{Distribution theory and Monte Carlo studies} \and
      {05.65.+b}{Self-organized systems}
     } 
} 
\maketitle
\section{Introduction}
\label{intro}
The first microscopic model of financial markets goes back to Bachelier in 1900 \cite{bachelier1900}, whose work is based on the hypothesis of independent variations modifying the value of the prices. He obtained a price that follows a random-walk, a disappointing result for those trying to make predictions on financial markets. However, there is now substantial empirical evidence that shows that price variations do not have a random-walk behaviour. In particular, the distribution of returns $P (r)$, a return being the relative price change in a given time interval, has been the subject of numerous empirical investigations. Instead of the Gaussian distribution expected from Bachelier model, different authors suggested that $P (r)$ behaves as an exponential \cite{mantegna95,bouchaud} or a power-law \cite{devries94,pagan96,guillaume97,gopi98,lux99,gopi99} for large values of $r$. No concensus has been reached for the exact expression of $P (r)$, if there is one, but it is now widely accepted that agents are not making decisions independently. Even if the panel of choices for investment opportunities is nearly infinite, agents tend to react to some common information, or at least, agents can be grouped by clusters of agents sharing the same information. The existence of these herds of agents is the basis of the model introduced by Egu\'{\i}luz and Zimmermann \cite{eguiluz99} (EZ model), which is a dynamical version of a previous model by Cont and Bouchaud \cite{cont97}. In the next section, we briefly describe the EZ model and detail its stationary solution obtained in \cite{dhulst-ijtaf}. We show numerically that a finite size effect with most of the agents buying or selling simultaneoulsy is present, a mechanism similar to a crash \cite{eguiluz99}.

An important parameter in the EZ model is the rate of investment $a$, which represents the probability of making an investment, $1- a$ being the probability that instead information propagates. In any real situation, we expect $a$ to be small and it can be shown that in the limit $a \rightarrow 0$, the model is in a critical state, with groups of investors of all sizes for infinite systems. In \cite{dhulst-physa}, a mechanism driving the system towards this critical state is proposed, based on a democratic decision process where every agent in a group takes part in the decision process. It is also shown that a dictatorship decision process with one agent making decision for a whole group leads to less volatile markets. These particular extensions of the model are presented in Sec. \ref{sec:the democratic and dictatorship models}. In the same section, we introduce a mixed model, where the decision processes can either be dictatorship with a probability $b$ or democratic with a probability $1-b$. Numerically, it is shown that the properties of the model are democratic for $b$ less than 0.5. The average investment rate stays to a very low value. When $b$ increases from 0.5 to 1.0, a continuous transition from democratic to dictatorship is underlined, with the average fragmentation rate increasing from a value close to 0 to a value close to 0.5. For $b \approx 0.8$, the effects of both types of decision processes cancel each other.

In the original EZ model, the fragmentation rate $a$ is a constant with a value chosen and fixed during a simulation. A more general approach of the problem suggests that $a$ should be considered as a function of the size of the clusters of agents, which is the only difference between the groups of agents. In Sec. \ref{sec:the democratic and dictatorship models}, we investigate numerically the functional dependence of $a$ as generated by the democratic and dictatorship extensions of the EZ model.

\section{The EZ model}
\label{sec:the ez model}

The Egu\'{\i}luz and Zimmermann model \cite{eguiluz99} is one of the simplest models you could imagine for herding. Agents are grouped by clusters of agents sharing the same information, and there is no information available at the beginning of the simulation. At each time step, one agent is chosen at random. With a probability $a$, he decides that it is the right time to invest and triggers all the agents sharing his information to invest with him. The cluster this agent belongs to is then fragmented into independent agents having no information to share. With a probability $1 - a$, the chosen agent decides that he would like to know a bit more before making an investment. Another agent is chosen at random and the two agents share their information. Hence, the clusters of both agents coagulate to form a single larger cluster. In other words, at each time step a cluster is fragmented with probability $a$, or two clusters coagulate with a probability $1 - a$. The number of agents, $N_0$, is fixed. 

The investment decision can either be buy or sell with equal probability. No feedback according to previous decisions has been implemented in the basic model, but it is quite easy to devise three choices models, buy, sell or share information, with memory \cite{dhulst-three}. When a cluster of agents decides to invest, it modifies the demand and supply equilibrium, which in turn affects the price of the exchanged commodity. As already mentioned, the price return $r$ is the relative price variation on a given time interval. The return $R$ here is defined to be the relative number of agents buying or selling at a particular time, taken with its sign. By convention, if agents are buying, the return is positive and it is negative if agents are selling. Hence, if $n_s$ is the number of clusters of size $s$, the probability to have a return $R$ of size $s/N_0$ is given by $s n_s / N_0$. $R$ is related to the price change $r$, usually using a logarithmic variation like \cite{cont97}

\begin{equation}
\ln P (t) - \ln P (t-1) = \frac{R}{\lambda}
\label{eq:return logarithmic}
\end{equation} 
where $P (t)$ is the price at time $t$ and $\lambda$ the market liquidity. $\lambda$ expresses the sensivity of a price to modifications in the supply and demand process. Alternatively, some authors consider that \cite{zhang99}

\begin{equation}
P(t) - P(t-1) = \lambda \sqrt{R}.
\label{eq:return square root}
\end{equation}
In either case, the cluster size distribution $n_s$ describes the supply and demand variations and is the quantity of interest.

A master equation for the number $n_s (t)$ of clusters of size $s$ at time $t$ can be written as \cite{dhulst-ijtaf}

\begin{eqnarray}
\nonumber
{\partial n_s\over \partial t} &=& -a s n_s + {(1-a)\over N_0} \sum_{r=1}^{s-1} r n_r (s-r) n_{s-r}\\
&& - {2(1-a) s n_s\over N_0} \sum_{r=1}^{\infty} r n_r\\
{\partial n_1\over \partial t} &=& a \sum_{r=2}^{\infty} r^2 n_r - {2(1-a)  n_1\over N_0} \sum_{r=1}^{\infty} r n_r.
\label{eq:time variation n1}
\end{eqnarray} 
Note that one time step in the continuous description is chosen to correspond to one attempted update per agent in the numerical simulation. In the first equation, the first term on the right hand side describes the fragmentation of a cluster of size $s$, the second term, the coagulation of two clusters to form a cluster of size $s$ and the last term, the coagulation of a cluster of size $s$ with another cluster. The second equation is the equation for the clusters of size one, with the first term on the right hand side describing the fragmentation of any clusters, which creates clusters of size one, while the second term is the coagulation of a cluster of size one with another cluster. 

\begin{figure}
\includegraphics[width=0.5\textwidth]{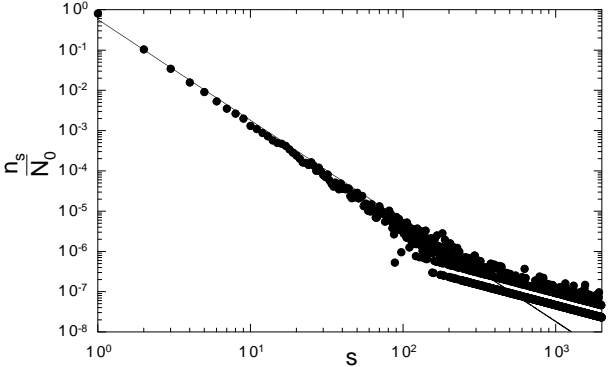}
\caption{Cluster size distribution $n_s / N_0$ for the EZ model for $N_0 = 10^4$ ($\bullet$) agents investing with a probability $a=0.01$, after $t = 10^6$ time steps. The continuous line is a guide to the eyes for a power law of exponent $\tau = 5/2$.}
\label{fig:cluster size distribution} 
\end{figure}

\begin{figure}
\includegraphics[width=0.5\textwidth]{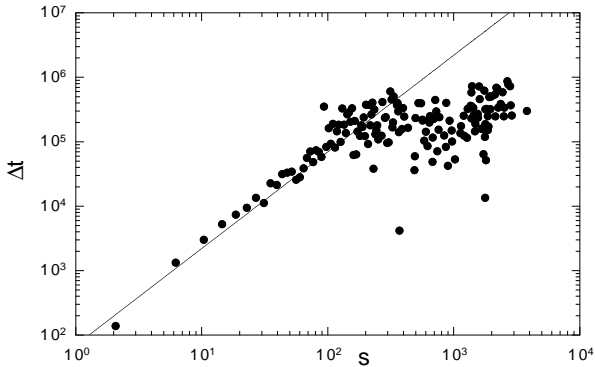}
\caption{Time difference $\Delta t$ between two events of size $s$, as a function of $s$. The simulation ran for $t=10^6$ time steps, with $N_0 = 10^4$ agents and $a=0.01$. The continuous line is a guide to the eye for a power-law of exponent $3/2$. Note that for $s>200$, large events are clearly below the continuous line.}
\label{fig:crash frequency} 
\end{figure}

The previous set of equations can be solved \cite{dhulst-ijtaf} to obtain a size distribution

\begin{equation}
\frac{n_s}{N_0} \sim \left({4 (1-a)\over (2-a)^2}\right)^s s^{-5/2}.
\label{eq:size distribution}
\end{equation}
The model displays a power-law distribution $n_s \sim s^{-\tau}$ of exponent $\tau = 5/2$, with an exponential cut-off. The exponential correction vanishes in the limit $a \rightarrow 0$. All the other stationary properties of the model can be calculated, like the moments of the distribution \cite{dhulst-ijtaf} or the connectivity $c$, defined as the average number of links per agent. The time average $\overline{c}$ of the connectivity, for instance, is equal to

\begin{equation}
\overline{c} = 2 \left( \frac{2-a}{1-a} \ln (2-a) - 1 \right).
\end{equation}
This result is obtained using the fact that a cluster of size $s$ always has $s-1$ links, or $2 (s-1)$ links per agent. In Fig. 
\ref{fig:cluster size distribution}, we present the cluster size distribution $n_s$ obtained for more than 50 numerical simulations of $10^6$ time steps each, for a system with $N_0 = 10^4$ agents and a fragmentation rate $a=0.01$. 
 The continuous line is a guide to the eye for a power-law of exponent $\tau = 5/2$. Taking the limit $a \rightarrow 0$ generates a time scale separation between the very quick propagation of information and the very slow process of decision making. It allows the existence of clusters of all sizes in infinite systems. However, for any finite size system, clusters of the order of the system size can be created when $a$ is tuned to a low value. This corresponds to the formation of a bubble, and when this cluster of agents decides to invest, it modifies drastically the supply and demand process, like when a bubble bursts. This can be associated with a crash phenomenon \cite{eguiluz99}, and corresponds to the events for large $s$ in the data presented in Fig. 
\ref{fig:cluster size distribution}. Note that for the choice of parameters, the largest events involve clusters of order $4.10^3$ agents, not shown in Fig. 
\ref{fig:cluster size distribution}. No precursory pattern characteristic of real crashes is expected here \cite{sornette96}, as the cooperative, or herding, behaviour is strictly localized in time. As soon as a cluster invests, this cluster is fragmented into independent agents. Hence, this cluster cannot be spotted before he invests, neither can it be the source of further cooperative behaviour. To illustrate the high frequency of large events, Fig. \ref{fig:crash frequency} shows the relation between the time interval $\Delta t$ between two events of size $s$ as a function of the size of these events, for the same choices of the parameters. 
The continuous line is a guide to the eye for a power-law of exponent $3/2$, which is the expected relation if the system was infinite. As can be infered from Fig. \ref{fig:crash frequency}, the time interval between two events of size $4.10^3$ should be larger than $10^7$ simulation time steps, ten times the actual length of the simulations. 

The relation between the value $\tau = 5/2$ for the exponent of the cluster size distribution and the value of the exponent for $P (r)$, the price return distribution is not straightforward. It is empirically found that $P (r) \sim 1/r^{\beta}$ for large $r$, with an exponent $\beta$ in the range 2 to 4 \cite{devries94,pagan96,guillaume97,gopi98,lux99,gopi99}. The latest estimates favour a value of $\beta$ close to 4, which means that neither of the proposed relations (\ref{eq:return logarithmic}) and (\ref{eq:return square root}) are convenient. This could be due to a wrong assumption for the relation linking the price return and the demand and supply equilibrium, but more likely, it is related to the simplicity of our model. For instance, agents could be acting similarly because they use the same broker and, unless they are loosing a lot of money, they will not change broker after each transaction. Hence, an improvment of the model could be to change the fragmentation or aggregation process. One extension that we considered was that $m$ agents could exchange information at each time step, instead of just two. So, at each time step, with a probability $a$, one cluster is fragmented and with a probability $1-a$, $m$ clusters coagulate. However, it was shown that the exponent of the power-law distribution is not affected by such a change \cite{dhulst-ijtaf}. 

The EZ model is a function of only two parameters, the number of agents $N_0$ and the investment rate $a$. Keeping $N_0$ fixed is unrealistic as the number of investors on financial markets is increasing. In real situations, we expect the change in the number of agents to be much slower than the trading rate so that the previous model can be considered in its stationary state with an increasing value of $N_0$ as time goes on. If the changes in the number of agents and the trading activities were on the same time scales, we should add a source term in Eq. (\ref{eq:time variation n1}). This would however prevent the system from reaching a stationary state, and time dependent equations would have to be considered. A time-dependent solution is unfortunately still not available. 

For the parameter $a$, it is assumed in the basic model that the investment rate $a$ is a constant. As agents are chosen at random, it suggests that bigger groups of agents have a higher rate of investment, because they are selected more often, that is, the effective investment rate is $a s$ rather than $a$. An interesting generalization of the model would allow $a$ to vary, and as the only difference between the groups of investors is their sizes, $a$ should be a function of the cluster size $s$ only. Work along these lines is in progress. A possible functional dependence of $a$ is discussed at the end of the next section.

\section{The democratic and dictatorship models}
\label{sec:the democratic and dictatorship models}

The previous model is very simple, with only two parameters, the number of agents $N_0$ and the fragmentation rate $a$. In the previous section, we discussed possible variations around these two parameters, allowing $N_0$ to vary or taking $a$ as a function of $s$, the size of the cluster of agents. However, if the origin of $N_0$ is clear, it is interesting to consider the mechanism that could generate a global parameter like $a$. In this section, we consider a model where each agent is given a microscopic parameter $p_i$, that represents her individuality. Agents interact with each other according to their $p_i$'s, whose value can change after interaction. We will show that this allows us to generate a macroscopic parameter like $a$. For simplicity, we consider that each $p_i$ is a random number chosen from a uniform distribution between 0 and 1.

\begin{figure*}
\includegraphics[width=0.5\textwidth]{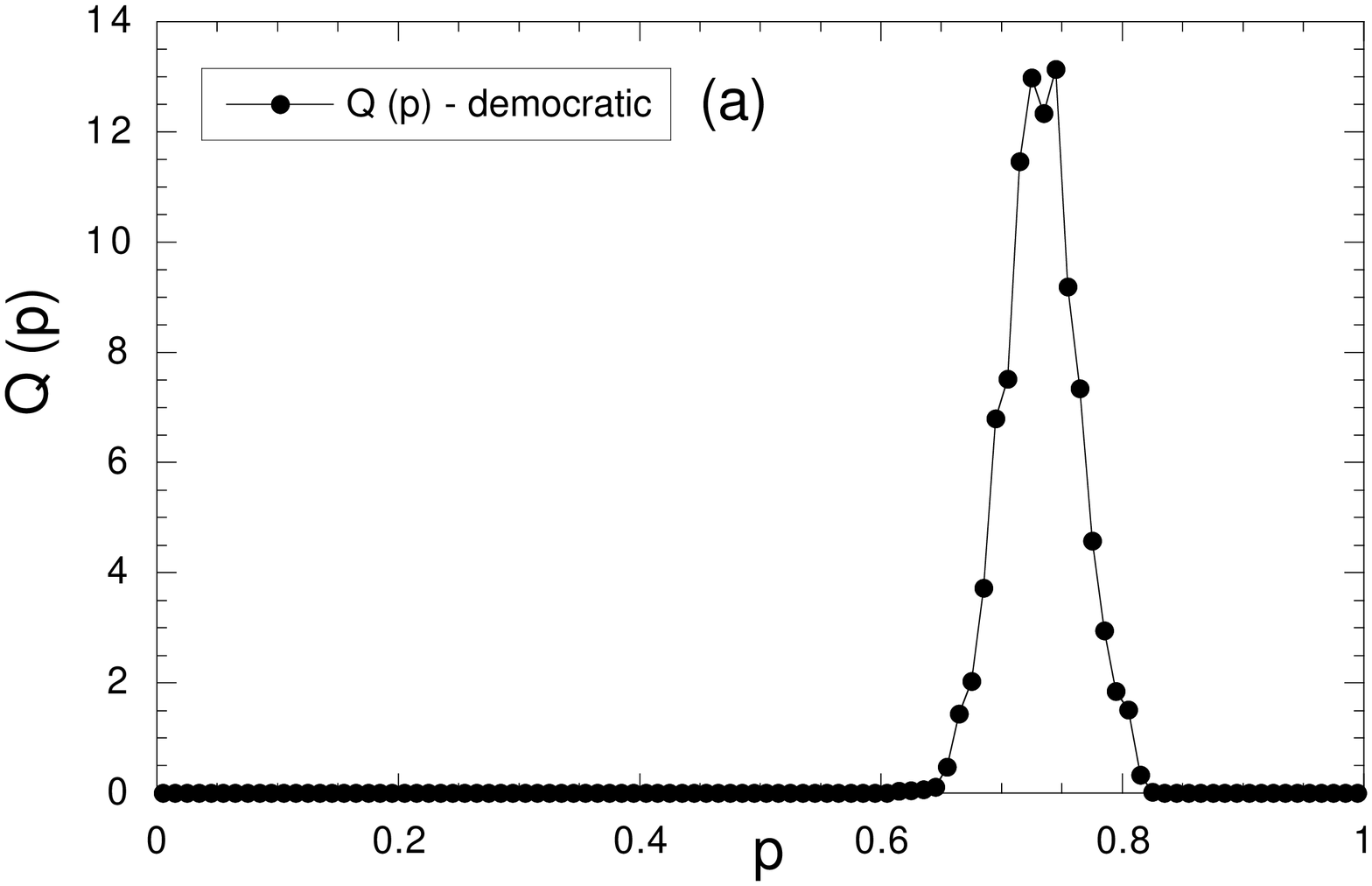}
\includegraphics[width=0.5\textwidth]{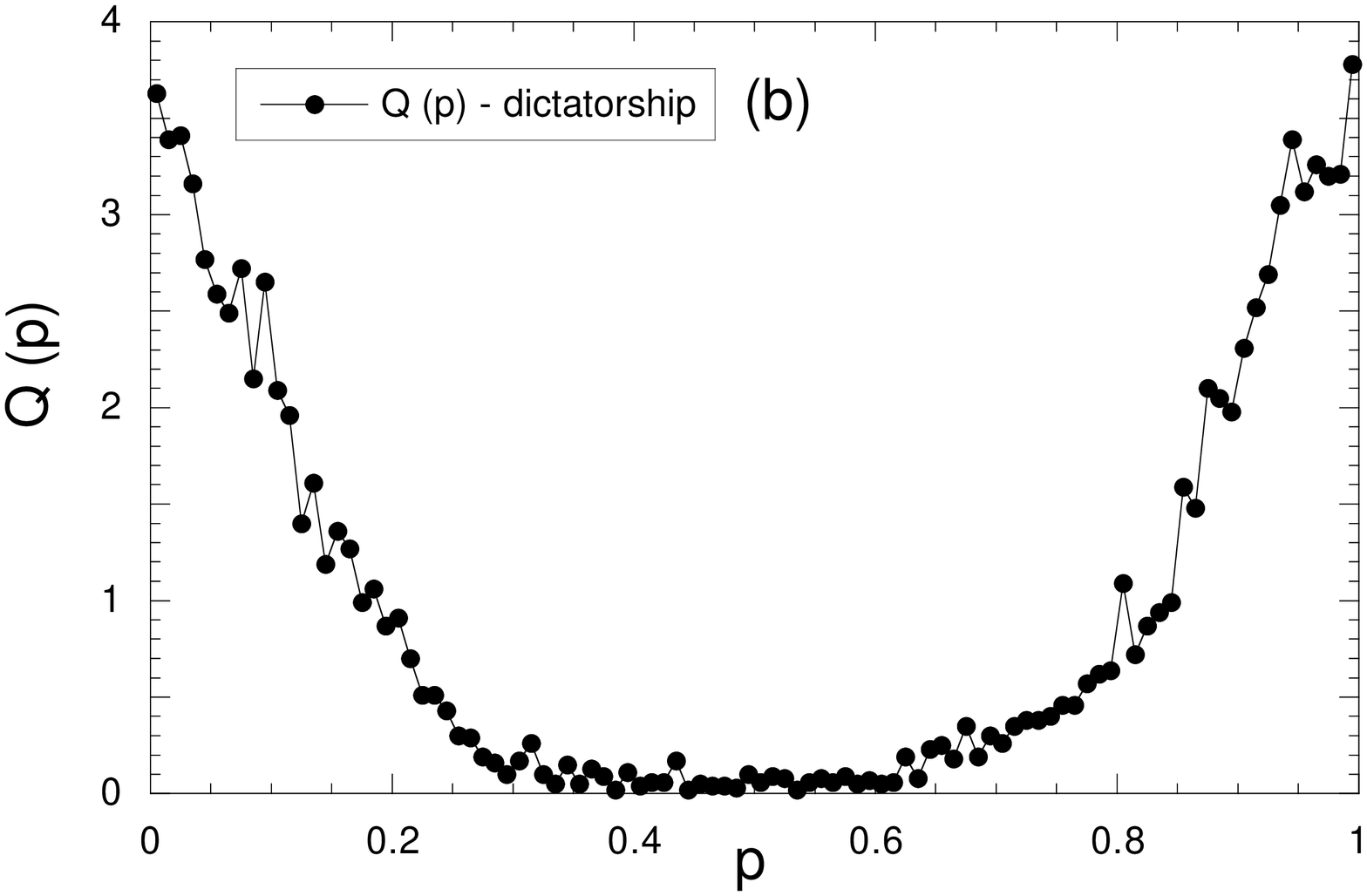}
\caption{Probability distribution $Q (p)$ of the characters $p$ of the agents for (a) the democratic and (b) the dictatorship model for $N_0 = 10^4$ agents, after $t=10^6$ time steps. The range is $R = 0.1$.}
\label{fig:dem and dict character distribution}
\end{figure*}
The principle of the model is very similar to the principle of the EZ model. Agents are grouped by clusters of agents sharing the same value of $p_i$. At each time step, two agents $i$ and $j$, with associated numbers $p_i$ and $p_j$ respectively, are selected at random. With a probability $a_{ij} = | p_i - p_j |$, agents $i$ decides to invest and triggers the action of his cluster. That is, all these agents are given new random number from the range $\lbrack p_i - R, p_i + R \rbrack$, where $R$ is a fixed number between 0 and 0.5. It corresponds to a synchronised investment made by people sharing the same information, a purchase or a sale with equal probability. With a probability $1 - a_{ij}$, $i$ and $j$ exchange their information, and their clusters coagulate. All agents belonging to the clusters of $i$ and $j$ are given the same new number $p_{ij}$. We consider two variations of the model, a democratic version where $p_{ij} = (p_i + p_j)/2$ and a dictatorship version where $p_{ij} = p_i$.

We identify the parameter $p_i$ to the character of an agent, or rather to the way an agent is perceived or is perceiving the market. Agents with similar values of $p_i$'s are more likely to be making similar decisions, so that when they meet, the probability that they decide to exchange information is high. According to their history of aggregation and fragmentation, agents are learning, and they receive a value of $p_i$ which is close to the value they had when they were in a cluster. The two different processes of aggregation refer to two different type of decision processes, an active one where each agent is taking part in the decision making, which we decide to call a democratic process, and a passive process where all the agents rely on one of them to make decision, which we call a dictatorship process. 

For both models we made numerical simulations to investigate the distributions $Q (p)$ of the $p_i$'s, that is, $Q (p) dp$ is the relative number of agents associated with a value of $p_i$ inside $(p, p+dp)$. The result for $Q (p)$ for the democratic model is presented in Fig. \ref{fig:dem and dict character distribution}a, for a system with $N_0 = 10^4$ agents and a range of $R = 0.1$. 
As the process of aggregation consists in averaging over the values of the $p_i$'s, the system is driven towards a coherent state where most of the agents have a value of $p_i$ spread around an average value $\overline{p}$, the amplitude of the spreading being related to the value of the range $R$. The exact value of $\overline{p}$ is meaningless as it originates from the averaging over the chosen initial distribution. Moreover, as the averaging is not equally weighted for all agents, it can be strongly history dependent \cite{dhulst-physa}. We define $\overline{a}$ to be the average value of $a_{ij}$ over $i$ and $j$. It corresponds to the macroscopic value $a$ of the EZ model. As most $p_i$'s are close to $\overline{p}$, $\overline{a}$ is close to 0 in the democratic model. Hence, the democratic model self-organizes into a coherent state where clusters of agents of all sizes exist, with a time scale separation between information transmission and investment. In Fig. \ref{fig:dem and dict character distribution}b, we present the results for the dictatorship model for $N_0 = 10^4$ agents and a range $R=0.1$. A spontaneous segregation into two equal sized populations happens, with one half of the agents associated with a value of $p_i$ around a value $p^{(1)}$ close to 0 and the other half associated with a value $p^{(2)}$ close to 1. The origin of this segregation is less obvious than the origin of a common value of $p$ for the democratic model. Due to this segregation into two populations, the value of $\overline{a}$ for the dictatorship model is approximately equal to 0.5. As a result, the dictatorship model displays a level of organization, but is not self-organized as the system is not in its critical state. It also means that the exponential cut-off of the size distribution present in Eq. (\ref{eq:size distribution}) is important and prevents large clusters of agents to develop. Hence, the dictatorship model generates less volatile markets than the democratic model.

\begin{figure*}
\includegraphics[width=0.5\textwidth]{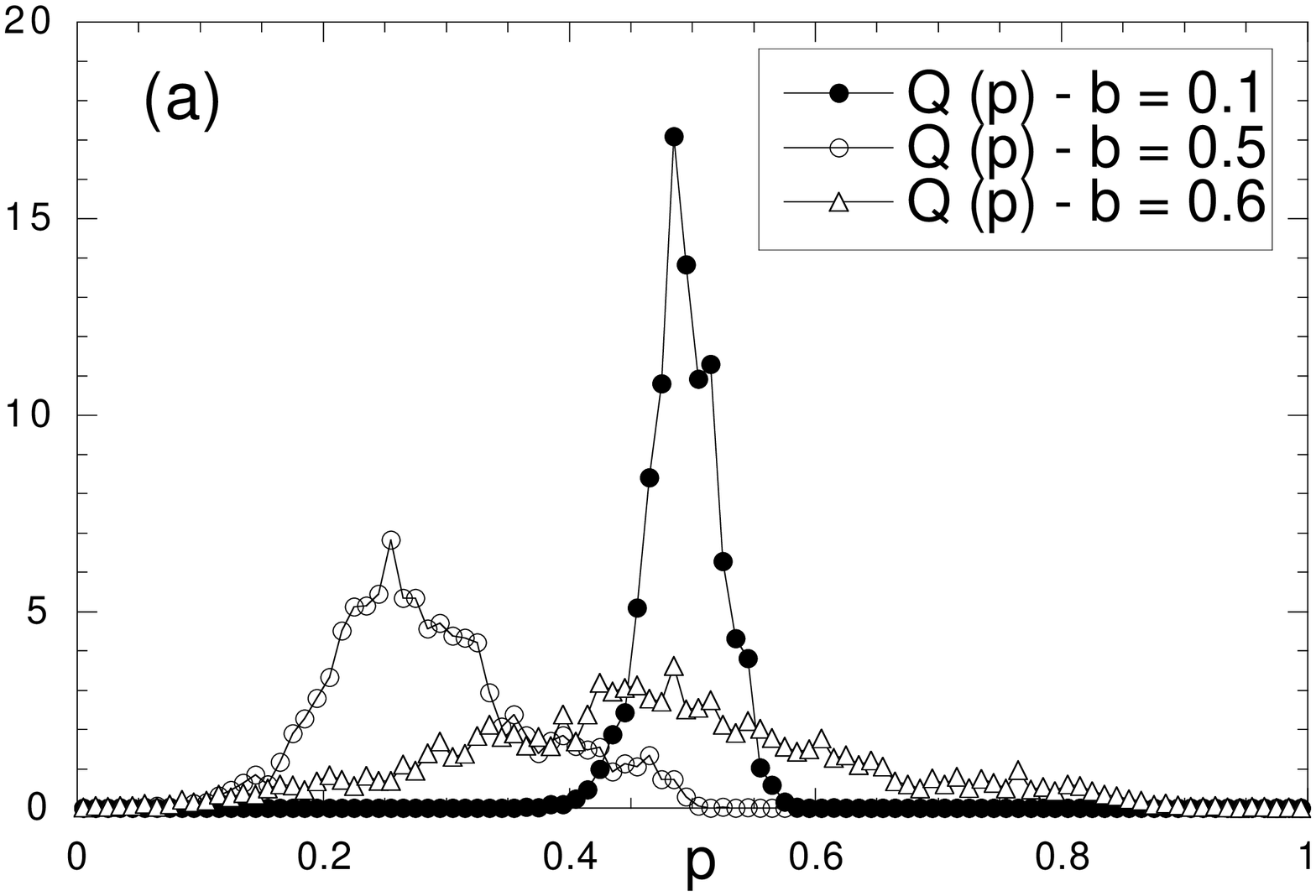}
\includegraphics[width=0.5\textwidth]{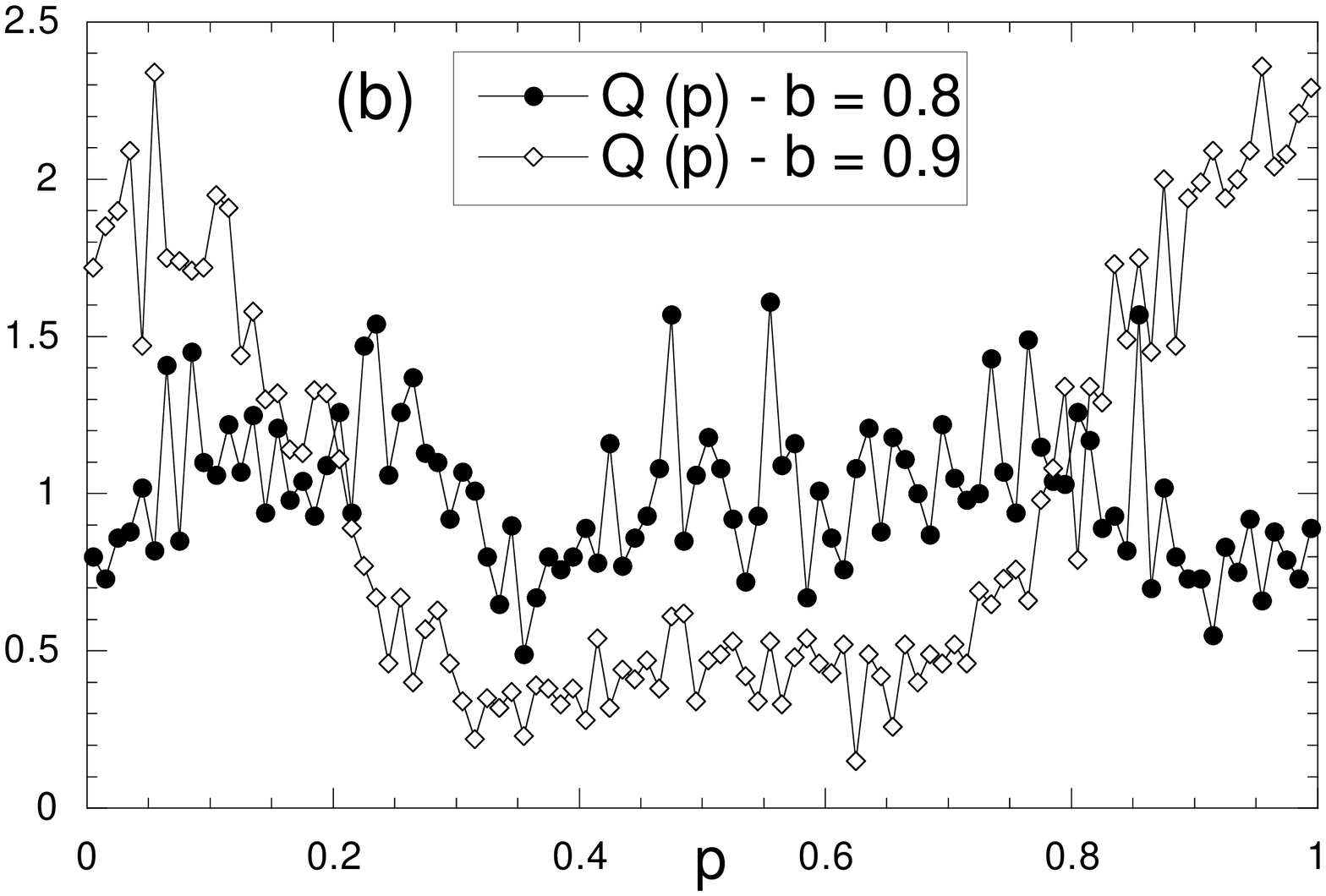}
\caption{Probability distribution $Q (p)$ of the characters $p$ of the agents for the mixed model for (a) $b=0.1$ ($\bullet$), 0.5 ($\circ$), 0.6 ($\triangle$) and (b) $b=0.8$ ($\bullet$), 0.9 ($\diamond$). All simulations where done for $N_0 = 10^4$ agents and lasted $t=10^6$ time steps. The range is $R = 0.1$. Simulations for $b$ less than 0.5 all give similar results to $b=0.1$.}
\label{fig:mixed character distribution}
\end{figure*}

A natural extension of the previous models consist in a mix of the democratic and dictatorship models, with a probability $b$ of having dictatorship associations, while democratic associations happen with a probability $1-b$. In this mixed version, some agents are passive, relying on another agent to make decisions for them, while others are active and make decisions for several agents. The resulting distributions $Q (p)$ of $p_i$'s are presented in Fig. \ref{fig:mixed character distribution}a for $b=0.1$, $b=0.5$ and $b=0.6$ and Fig. \ref{fig:mixed character distribution}b for $b=0.8$ and $b=0.9$. 
As a reminder, $b=0$ is the democratic model and was presented in Fig. \ref{fig:dem and dict character distribution}a, while $b=1$ is the dictatorship model, shown in Fig. \ref{fig:dem and dict character distribution}b. All simulations where performed over $10^6$ time steps, with $N_0 = 10^4$ agents, a range $R=0.1$ and a initial uniform distribution between 0 and 1 for $Q (p)$. The democratic aggregation process is dominant for all value of $b$ less than 0.5 as seen for $b=0.1$. So, for $b$ less than 0.5, the stable equilibrium of the system is a coherent state. From $b=0.5$, the competition between democratic and dictatorship associations tends to compensate each other, which flattens the distribution $Q (p)$. A flat distribution is achieved for $b \approx 0.8$, as seen in Fig. \ref{fig:mixed character distribution}b. Also in Fig. \ref{fig:mixed character distribution}b, it can be seen that for $b$ higher than 0.8, the dictatorship aggregation process is dominant, with a clear segregation of the population. 

\begin{figure}
\includegraphics[width=0.5\textwidth]{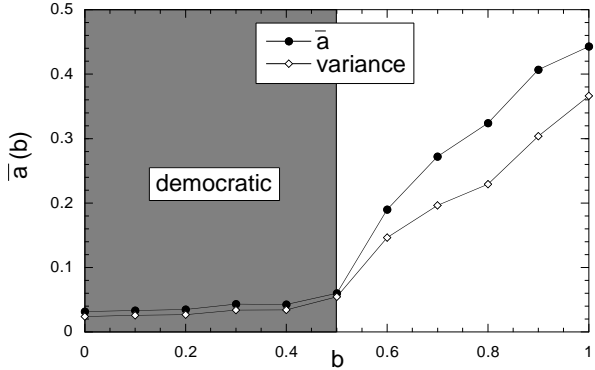}
\caption{Average fragmentation rate $\overline{a}$ ($\bullet$) as a function of $b$. For $b$ less than 0.5, the system converges towards a coherent state with $\overline{a}$ close to 0, which is denoted by the shaded area in the figure. The variance of $\overline{a}$ is also presented ($\diamond$) and is of the same order than $\overline{a}$. All simulations were performed during $10^6$ time steps for systems of $N_0 = 10^4$ agents and a range $R=0.1$.}
\label{fig:a as a function of b}
\end{figure}

Another way of examining the transition consists in measuring the distribution $N (| p_i - p_j |)$, which is defined to be the frequency of appearance of a given difference $| p_i - p_j |$. Remember that the average value of $| p_i - p_j |$ over all agents corresponds to the average fragmentation rate $\overline{a}$. By looking at the distribution $N (| p_i - p_j |)$, we investigate the microscopic structure of the fragmentation rate. This distribution is related to the distribution $Q (p)$ by
\begin{figure}
\includegraphics[width=0.5\textwidth]{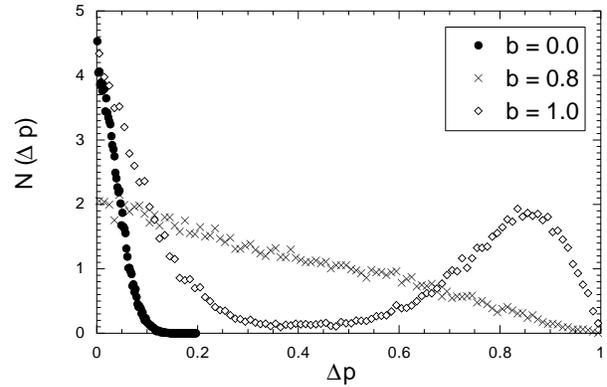}
\caption{Probability distribution $N (\Delta p)$ of selecting two agents at random with characters $p_i$ and $p_j$ such as $\Delta p = | p_i - p_j |$.  Each simulation is done for $N_0 = 10^4$ agents, $R=0.1$ and laster $t=10^6$ time steps. The choices for $b$ are $b=0$ (democratic, $\bullet$), $b=0.8$ ($\times$) and $b=1$ (dictatorship, $\diamond$). The results for $b=0$ have been divided by 5 for ease.}
\label{fig:fragmentation rate distribution}
\end{figure}

\begin{equation}
N (\Delta p) = \int_0^1 \int_0^1 dp_1 dp_2 Q (p_1) Q (p_2) \delta ( | p_1 - p_2 | - \Delta p ),
\end{equation}
where we use the notation $ | p_i - p_j | = \Delta p$. In Fig. \ref{fig:fragmentation rate distribution}, we present the distribution $N (\Delta p)$ for $b=0$, the democratic model, $b=0.8$ and $b=1.0$, the dictatorship model. The simulations were performed during $10^6$ time steps for $N_0 = 10^4$ and a range $R = 0.1$. 
Note that for ease of presentation, we have divided the results for $b=0$ by 5. From $b=0$ to $b\approx 0.8$, the distribution $N (\Delta p)$ is the positive half of a Gaussian distribution centered at the origin. The spread of this Gaussian stays constant for $b$ less than 0.5 and starts to increase with $b$ for $b\ge 0.5$. At around $b=0.8$, the distribution becomes a straight line, in agreement with a flat distribution for $Q (p)$. In this case, $N ( \Delta p) = 2 (1 - \Delta p)$. For $b$ larger than 0.8, $N (\Delta p)$ has two maxima, one at $\Delta p = 0$ and one at $\Delta p$ close to 1, as can be seen for $b=1.0$ in Fig. \ref{fig:fragmentation rate distribution}. The second maximum tends to 1 as $R$ is decreased towards 0. From these distributions, the average value of the fragmentation rate $\overline{a}$ can be infered and is presented in Fig. \ref{fig:a as a function of b} as a function of $b$. 
For $b<0.5$, $\overline{a}$ stays nearly constant and close to 0, going to 0 as the $R$ is decreased. This part of the parameter space is refered to as the democratic phase in Fig. \ref{fig:a as a function of b}, in agreement with the existence of a coherent state. At $b=0.5$, $\overline{a}$ increases towards a value close to 0.5 at $b=1$. Also presented in Fig. \ref{fig:a as a function of b} is the variance of $\overline{a}$. It is important to note that the variance of $\overline{a}$ is always of the order of $\overline{a}$ itself. For small values of $b$, this is because $\overline{a}$ and its variance are both of the order of $R$, while for $b$ close to 1, this is because the distribution $N (\Delta p)$ has to maxima, one in zero and one in one, not just one maximum in 0.5.

The transition outlined here, from a coherent state to a segregated population, is very similar to the transition investigated in \cite{zanette97}. The average fragmentation rate can be associated to the order parameter of the transition. Unlike Ref. \cite{zanette97}, our numerical results suggest that the transition is second-order, but further investigations are required. 

\begin{figure*}
\includegraphics[width=0.5\textwidth]{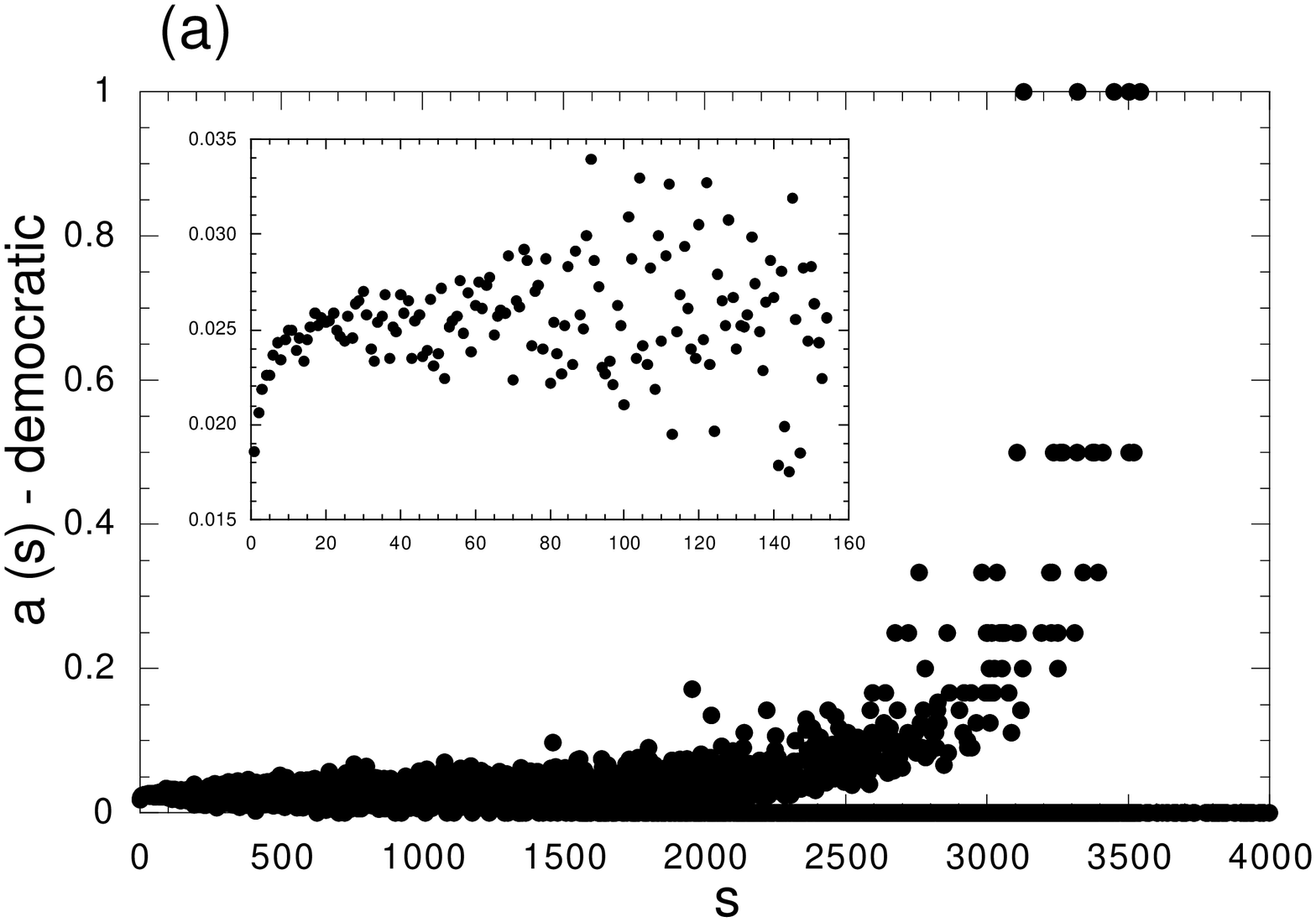}
\includegraphics[width=0.5\textwidth]{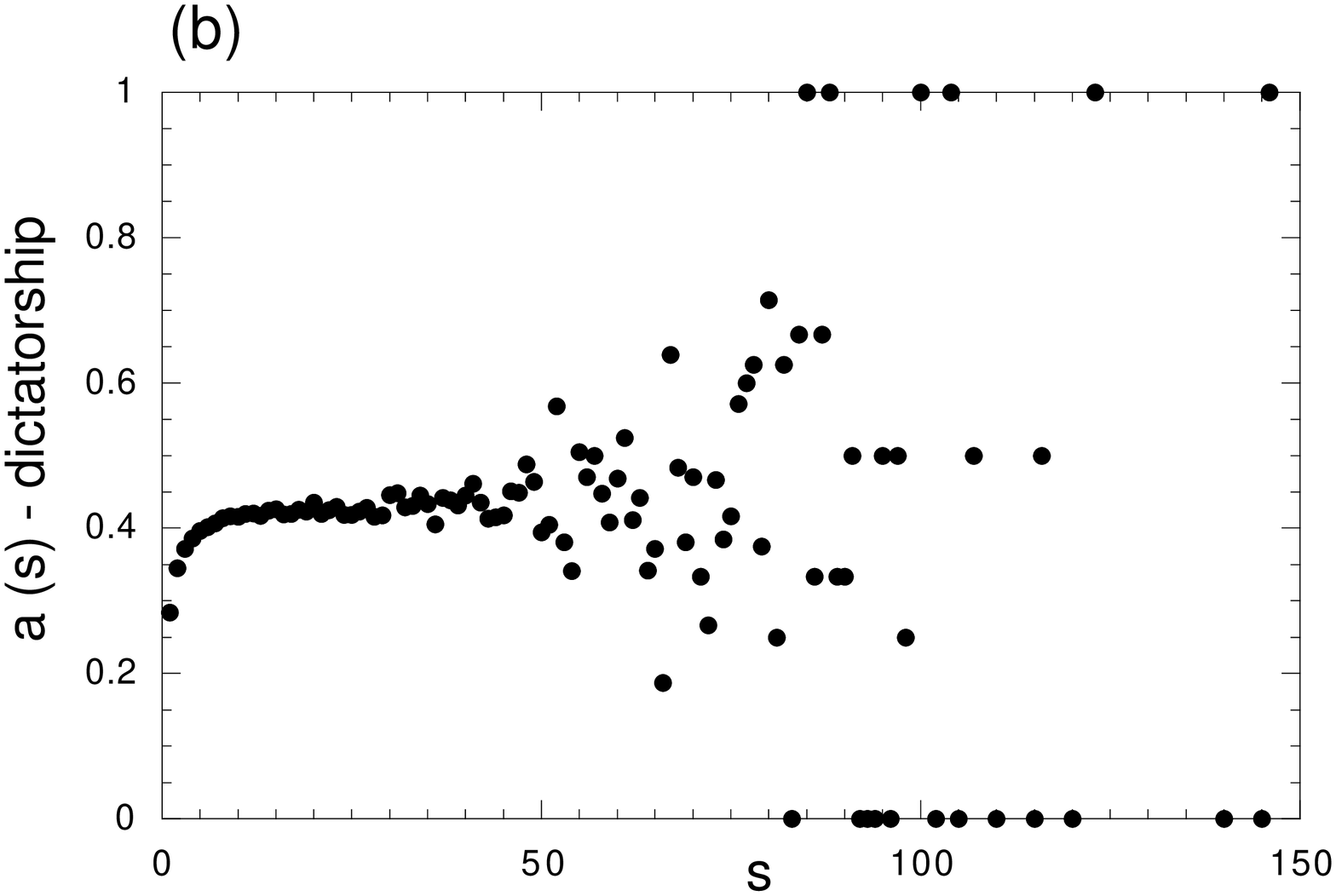}
\caption{Average fragmentation rates of clusters of size $s$ as a function of $s$ for (a) the democratic and (b) the dictatorship models. The inset in Fig. \ref{fig:a as a function of s}a allows us to appreciate the behaviour of $a$ for low values of $s$. We simulated systems of $N_0 = 10^4$ agents with $R=0.1$ during $10^7$ time steps.}
\label{fig:a as a function of s}
\end{figure*}

As discussed at the end of the previous section, the EZ model is arbitrary in its functional choice for $a$, taking it to be a simple constant. On the contrary, in the democratic and dictatorship models, a macroscopic parameter $\overline{a}$ is generated by the interaction between agents. We investigate numerically the parameter $\overline{a}$ of the democratic and dictatorship models, seen as a function of the cluster sizes, to question the functional choice of the EZ model. The variation of $\overline{a} (s)$ in the democratic model is presented in Fig. \ref{fig:a as a function of s}a for $N_0 = 10^4$ agents and a range $R=0.1$. The simulation ran for $10^7$ time steps. It can be seen that $\overline{a} (s)$ stays very low for $s$ less than around 2000, while the larger events have just happened one or two times. Hence, they are not statistically relevant, but it can be seen that most of the large clusters selected have not broken down, which implies a very low value for $\overline{a}$. The inset of Fig. \ref{fig:a as a function of s}a presents a zoom of the figure for the lowest sizes and it can be seen that for $s$ less than around 20, the fragmentation rate is particularly low. $\overline{a}$ then stabilizes at a value around 0.025, with large fluctuations that are increasing as $s$ is increased. 
In Fig. \ref{fig:a as a function of s}b is presented $\overline{a}$ as a function of $s$ for the dictatorship model for a simulation of $10^7$ time steps, with $N_0 = 10^4$ agents and a range $R=0.1$. For $s$ less than 10, $\overline{a}$ is very low, while stabilizing around $\overline{a} \approx 0.42$ from $s\approx 10$ to $s \approx 50$. For $s > 50$, less and less event are recorded, leading to huge fluctuations of $\overline{a}$. As a conclusion, if the democratic or the dictatorship aggregation processes are representative of the mechanism that generates a macroscopic parameter $a$ in the EZ model, taking $a$ as a constant seems to be a good approximation. A more elaborate model should consider that $a$ starts from a very low value for $s=1$, then reaches a stationary value for small $s$ and starts to fluctuate around this value as $s$ increases, the fluctuations also increasing with $s$.    

\section{Conclusions}
\label{sec:conclusions}

The EZ model has been presented as a simple model to mimic the competition between information transmission and decision making in financial markets. Due to shared information, agents do not act independently and make group decisions, a phenomenon known as herding. These group decisions can have a strong impact on the market, modifying drastically the supply and demand equilibrium and, ultimately, the price. The exact stationary solution of the EZ model is obtained in the limit of infinite size systems. When there is a time scale separation between the quick information transmission and the slow decision making, the model is in its critical state, with groups of agents of all sizes. The size distribution of the groups of agents is a power-law of exponent $5/2$, with an exponential cut-off. For finite size systems close to the critical state, clusters of agents can merge to induce large events, a phenomenon similar to a crash. We show numerically that these events happen far more often than expected from the solution for infinite size systems. However, due to the dynamics of the system, no crash precursor patterns can be identified. 

The previous model suffers from a dependence on a global parameter $a$ which represents the investment rate. This parameter is supposed to be the same for all agents, which is very unlikely, and is externally controled, another unrealistic feature. Two extensions of the EZ model are presented, a democratic version where all agents take part in the decision process, and a dictatorship version, where one agent is making the decision for several others. Both extensions display a level of organization, with the democratic model driving the system into its critical state, while the dictatorship model displays a spontaneous segregation in the population of agents. The average investment rate is close to 0 in the democratic model, close to 0.5 in the dictatorship model. We introduced in this paper a mixed version, with a probability $b$ of a dictatorship decision, and $1-b$ of a democratic decision. This allows us to induce a transition from a coherent state, which corresponds to the democratic model, to a bistable state, the dictatorship model. The transition, which happens for $b$ close to 0.5, seems to be second-order, with the average investment rate as the order parameter. 

Another arbitrary feature of the EZ model is that the investment rate $a$ is taken as a constant, leading to a higher investment rate for the larger groups of agents because they are selected more often. It seems reasonable to allow this investment rate to depend on the size of the groups of agents. As we proposed the democratic and dictatorship models to simulate the demand process, we have investigated numerically the variation of the average investment rate $\overline{a}$ as a function of the size $s$ of the groups of agents in these models. We showed that $\overline{a}$ stays nearly constant for all values of $s$, but with huge fluctuations as $s$ increases. Also, for small values of $s$, $\overline{a}$ decreases towards 0. We argue that taking the investment rate as a constant in the EZ model is a good approximation, if the democratic or dictatorship mechanisms are representative of the demand process.

\end{document}